\documentclass[prb,aps,showpacs,showkeys]{revtex4}
\usepackage{amsmath, amssymb, graphicx, subfigure}
\usepackage{graphicx}
\usepackage{epsfig}
\usepackage{color}
\begin{document}
\newcommand{\hh}{\hspace{0.2in}}
\newcommand{\vv}{\vspace{0.2in}}
\newcommand{\mbb}{\mathbf}
\newcommand{\ex}{\texttt{e}}
\newcommand{\mc}{\mathcal}
\def\tensor{\stackrel{\leftrightarrow}}

\title{Effects of periodic scattering potential on Landau quantization and ballistic transport of electrons in graphene}

\author{Godfrey Gumbs$^{1,2}$, Andrii Iurov$^{1}$, Danhong Huang$^{3}$, Paula Fekete$^{4}$ and Liubov Zhemchuzhna$^{5}$}

\affiliation{
$^{1}$ Department of Physics and Astronomy Hunter College of the City University of New York \\
695 Park Avenue, New York, NY 10065 \\
$^{2}$ Donostia International Physics Center (DIPC), Manuel Lardizabal \\
4, E-20018, San Sebastian, Basque Country, Spain  \\
$^{3}$ Air Force Research Laboratory, Space Vehicles Directorate, \\
Kirtland Air Force Base, Kirtland, NM \\
$^{4}$ West Point Military Academy, West Point, NY\\
$^{5}$ Department of Math and Physics, NCCU, Durham, NC 27707, USA.
}


\date{\today}

\begin{abstract}
     A two-dimensional periodic array of scatterers has been introduced to a single layer
of graphene in the presence of an external magnetic field perpendicular to the graphene
layer. The eigenvalue  equation for such a system has been solved numerically to display
the structure of split Landau subbands as functions of both wave number and magnetic flux.
The effects of pseudo-spin coupling and Landau subbands mixing by a strong scattering potential
have been demonstrated. Additionally, we  investigated the square barrier tunneling problem
when magnetic field is  present, as well as demonstrate the crucial difference in the modulated
band structure between  graphene and the two-dimensional electron gas. The low-magnetic field
regime is particularly interesting for   Dirac fermions and  has been  discussed. Tunneling
of Dirac electrons through a magnetic potential barrier has been investigated to complement
the reported results on electrostatic potential scattering in the presence of an ambient magnetic field.
\end{abstract}


\maketitle


Dating back to the early work of Azbel \cite{azbel} and Hofstadter\cite{hofstadter},
the single-particle spectrum of a two-dimensional structure in the presence of both
a periodic potential and a uniform   ambient  perpendicular magnetic field has captivated
researchers for many years \cite{GG}$^{-}$\cite{cui1}. The  paper by   Hofstadter \cite{hofstadter} was 
for the energy spectrum of a periodic square lattice in the tight-binding approximation and subject 
to a perpendicular magnetic field. Ever since that time, there have been complementary calculations
for the hexagonal lattice \cite{GG},the two-dimensional electron gas (2DEG) with an electrostatic
periodic modulation potential \cite{pfannkuche,pfannkuche2,cui1} and even bilayer graphene where 
different stacking of the two types of atoms forming the sublattices  was considered \cite{bilayer}.
It has been claimed that one may be able to  observe evidence of the existence of Hofstadter's butterfly
in such experimentally measured quantities as density-of-states and conductivity \cite{ye, nat1, nat2}.
\par
The challenge facing experimentalists so far has been to carry out experiments on 2D structures
at achievable magnetic fields where the Hofstadter  butterfly  spectrum is predicted.  One
may follow the calculation of Hofstadter by using Harper's  equation
which may be regarded as a tight-binding approximation of the Schrodinger equation.  Then,
assuming that the magnetic flux through unit cell of the periodic lattice is a rational fraction
$p/q$ of the flux quantum in conjunction with the Bloch condition for the  wave function, one obtains
a $p\times p$ Hamiltonian matrix to determine the energy eigenvalues since one only  needs to
solve the problem in a unit cell.    Hofstadter himself was concerned  about ever
reaching magnetic fields where the rich self-similar structure of the butterfly  would be experimentally
observed due to the estimated high fields required to achieve this.
\par
In this paper, we propose  applying an electrostatic modulation potential to a flat sheet of graphene
in the presence of a perpendicular magnetic field to  produce the energy spectrum with self-similarity
at reasonably low magnetic fields. For a review, see \cite{sokoloff}. We also compare our results with those for
a modulated two-dimensional electron gas and discuss the difference. For completeness, we briefly
review tunneling of Dirac electrons through a magnetic barrier.

\par
\medskip
\section{Model and Theory}


In the presence of a periodic two-dimensional scattering potential modulation\,\cite{cui1}

\begin{equation}
V(x,\,y)=V_0\left[\cos\left(\frac{2\pi x}{d_x}\right)\cos\left(\frac{2\pi y}{d_y}\right)\right]^{2N}\ ,
\end{equation}
where $N=1,\,2,\,\cdots$ is an integer, $V_0$ is the modulation amplitude, and $d_x,\,d_y$
are the modulation periods in the $x$ and $y$ directions, respectively, we write the Hamiltonian
operator as
\begin{equation}
\hat{\mc H} = v_F \left[{
\begin{array}{cccc}
V(x,\,y)  &  {\hat{p}_x+eB_0y\hat{x}_0  + i\hat{p}_y } & 0 & 0  \\
{\hat{p}_x+eB_0y\hat{x}_0 - i\hat{p}_y } & V(x,\,y) & 0 & 0 \\
0 & 0 & V(x,\,y) & {\hat{p}_x+eB_0\hat{x}_0 - i\hat{p}_y } \\
0 & 0 & {\hat{p}_x+eB_0\hat{x}_0 + i\hat{p}_y } & V(x,\,y) 
\end{array}
}\right]
\end{equation}
\medskip
\begin{equation}
\hat{\cal H} = v_F \left[{
\begin{array}{cc}
\hat{\cal H}_K+V(x,\,y)\hat{I}  &  0 \\
0 & \hat{\cal H}_{K'}+V(x,\,y)\hat{I}  
\end{array}
}\right]\ .
\end{equation}
Here, $v_F$ is the Fermi velocity, $\hat{I}$ is a $2\times 2$ identity matrix. In this  system,
the magnetic flux through  unit cell is $\Phi=B_0(d_xd_y)$, which is assumed to be a rational
multiple of the flux quantum $\Phi_0=h/e$, i.e., $\beta\equiv\Phi/\Phi_0=p/q$ is an
irreducible fraction and  $p$ and $q$  are   integers. Furthermore, we choose the first Brillouin
zone defined by $ \vert k_x \vert \leq \pi/d_x$ and $|k_y|\leq \pi/(qd_y)$. By using the Bloch-Peierls condition,
the wave function of this modulated system may be expanded as

\begin{equation}
\Phi^{\pm}_{\ell;\,n,{\vec k}_{||}} \left( {x,\,y} \right)=\frac{1}{\sqrt{2{\cal N}_y}}\sum\limits_{s=-\infty}^{\infty}\left\{{\rm e}^{ik_y\ell_B^2(sp+\ell)K_1}\right.
\times\left.\left[\Psi_{n,k_x-(sp+\ell)K_1}^{K,\pm} \left( {x,\,y} \right)+\Psi_{n,k_x-(sp+\ell)K_1}^{K',\pm} \left( {x,\,y} \right)\right]\right\}\ ,
\end{equation}
where ${\vec k}_{||}=(k_x,\,k_y)$, ${\cal N}_y=L_y/(qd_y)$ is the number of unit cells, which
are spanned by $b_1=(d_x,\,0)$ and $b_2=(0,\,qd_y)$, in the $y$ direction, $L_y$ ($\to\infty$) is the sample
length in the $y$ direction, $K_1=2\pi/d_x$ is the reciprocal lattice vector in the $x$ direction and $\ell=1,\,2,\,\cdots,\,p$ is a new quantum number for labeling split $p$ subbands from a
$k_x$-degenerated landau level in the absence of modulation.
The above wave function satisfies the usual Bloch condition, i.e.,
\begin{equation}
\Phi^{\pm}_{\ell;\,n,{\vec k}_{||}} \left( {x+d_x,\,y+qd_y} \right)={\rm e}^{ik_xd_x}\,{\rm e}^{ik_yqd_y}\,\Phi^{\pm}_{\ell;\,n,k_{||}} \left( {x,\,y}\right)\ .
\end{equation}
Since the wave functions at the $K$ and $K^\prime$ points are decoupled from each other for
single-layer graphene, which is different from  bilayer graphene  \cite{petters},
we may write out explicitly the full expression for the wave function at these two points, i.e.,

\[
\Phi^{K,\pm}_{\ell;\,n,{\vec k}_{||}} \left( {x,\,y} \right)=\frac{C_n}{\sqrt{{\cal N}_yL_x}}\sum\limits_{s=-\infty}^{\infty} {\rm e}^{i[k_x+K-(sp+\ell)K_1] x} 
{\rm e}^{ik_y\ell_B^2(sp+\ell)K_1} \times 
\]
\vspace{0.1in}
\begin{equation}
\times
\left[{
\begin{array}{c}
\alpha_n^{\pm}\phi_{n-1,k_x+K-(sp+\ell)K_1} \left( y \right) \\
\phi_{n,k_x+K-(sp+\ell)K_1} \left(y \right) \\
0 \\
0
\end{array}
}\right] \ ,
\end{equation}
\[
\Phi^{K',\pm}_{\ell;\,n,{\vec k}_{||}} \left( {x,\,y} \right)=\frac{C_n}{\sqrt{{\cal N}_yL_x}}\sum\limits_{s=-\infty}^{\infty}
{\rm e}^{ik_y\ell_B^2(sp+\ell)K_1} \times{\rm e}^{i[k_x+K'-(sp+\ell)K_1] x} \times
\]
\begin{equation}
\times
\left[{
\begin{array}{c}
0 \\
0 \\
\phi_{n,k_x+K'-(sp+\ell)K_1} \left( y \right)  \\
\alpha_n^{\pm}\phi_{n-1, k_x+K'-(sp+\ell)K_1} \left( y \right)
\end{array}
}\right]\ .
\end{equation}
From here onwards, we will concentrate on the $K$ point by setting $K=0$ for convenience
 and omit the label for the $K$ point thereafter, i.e.,
\begin{widetext}
\begin{eqnarray}
\nonumber 
&& \psi^{A,\pm}_{\ell,n,{\vec k}_{||}} \left( {x,\,y} \right)\equiv\langle {\vec r}_{||}|\ell,n,{\vec k}_{||};\,A,\pm\rangle
=\frac{\alpha_n^{\pm}C_n}{\sqrt{{\cal N}_yL_x}}\sum\limits_{s=-\infty}^{\infty}
{\rm e}^{ik_y\ell_B^2(sp+\ell)K_1}   \times{\rm e}^{i[k_x+K-(sp+\ell)K_1] x}\,\phi_{n-1,k_x+K-(sp+\ell)K_1} \left( y \right) \ , \\
\\
\nonumber
&&  \psi^{B,\pm}_{\ell,n,{\vec k}_{||}} \left( {x,\,y} \right)\equiv\langle {\vec r}_{||}|\ell,n,{\vec k}_{||};\,B,\pm\rangle
=\frac{C_n}{\sqrt{{\cal N}_y L_x}}\sum\limits_{s=-\infty}^{\infty}
{\rm e}^{ik_y\ell_B^2(sp+\ell)K_1} \times{\rm e}^{i[k_x+K-(sp+\ell)K_1] x}\,\phi_{n,k_x+K-(sp+\ell)K_1} \left( y \right) \ ,  \\
\end{eqnarray}
\end{widetext}
where ${\vec r}_{||}=(x,\,y)$. A tedious but straightforward calculation gives rise to an
explicit expression for the matrix elements for the potential $V(x,\,y)$ as
\begin{widetext}
\begin{eqnarray}
&& V^{\ell',n',\mu'}_{\ell,n,\mu}(\vec{k}_{||})\equiv\sum\limits_{{\vec k}'_{||}}
\int\int\,dxdy\,\left[\Phi^{\mu'}_{\ell';\,n',{\vec k}'_{||}} \left( {x,\,y} \right)\right]^\dag\,V(x,\,y)\,\Phi^{\mu}_{\ell;\,n,{\vec k}_{||}} \left( {x,\,y} \right) \\
\nonumber
&&  =\frac{V_0C_{n'}C_n}{4^{2N}}\left\{{\rm e}^{ik_y\ell_B^2K_1(\ell-\ell')}\,\sum\limits_{i=0}^{N-1}\sum\limits_{j=0}^{N-1}\,\left[{\cal F}^{(B)}_{ij}+\alpha_{n'}^{\mu'}\alpha_n^{\mu}\,{\cal F}^{(A)}_{ij}\right]\right.
\left.+\delta_{\ell,\ell'}\delta_{n,n'}\,\left(1+\alpha_{n'}^{\mu'}\alpha_n^{\mu}\right)
\left[\frac{(2N)!}{(N!)^2}\right]^2\right\}\ , \\
\end{eqnarray}
where
\begin{eqnarray}
\nonumber
&& {\cal F}_{ij}^{(B,A)}=\left({
\begin{array}{c}
2 N \\
i
\end{array}
}\right)
\left({
\begin{array}{c}
2N \\
N
\end{array}
}\right)
\mc{T}_1
+
\left({
\begin{array}{c}
2N \\
j
\end{array}
}\right)
\left({
\begin{array}{c}
2N \\
N
\end{array}
}\right)\mc{T}_2 + 2 
\left({
\begin{array}{c}
2N \\
i 
\end{array}
}\right)
\left({
\begin{array}{c}
2 N \\
j
\end{array}
}\right) \mc{T}_3 \ , \\
\nonumber
&& \mc{T}_1 = A^{(B,A)}_1(0,\,N-i) \ , \\
\nonumber
&& \mc{T}_2 = A^{(B,A)}_2(N-j,\,0) \ , \\
\nonumber
&& \mc{T}_3 = A^{(B,A)}_3(N-j,\,N-i)  \ , 
\end{eqnarray}
\end{widetext}
as well as  the binomial expansion coefficient with $m \geq n$
\begin{equation}
\left({
\begin{array}{c}
m \\
n
\end{array}
}\right) \equiv \frac{m!}{n!\,(m-n)!} \ .
\end{equation}
In the above expressions, we have also introduced the following three self-defined functions

\begin{equation}
A_1^{(B,A)}(r,\,s)=D_{n',n}^{rs{(B,A)}}\,T_{\ell}^s\,\delta_{\ell,\ell'}\ ,
\end{equation}

\begin{eqnarray}
&& A^{(B,A)}_2(r,\,s)=D_{n',n}^{rs{(B,A)}}\left\{\delta_{\ell-\ell',r}\left[{\rm sgn}(n'-n)\right]^{\xi}+\delta_{\ell'-\ell,r}\left[{\rm sgn}(n-n')\right]^{\xi}\right\}\ , \\
\nonumber
&&  A^{(B,A)}_3(r,\,s)=D_{n',n}^{rs{(B,A)}}\left\{\delta_{\ell-\ell',r}\left[{\rm sgn}(n'-n)\right]^{\xi}\cos[\Theta_{rs}^{\ell'}(n',\,n)]\right. \left.+\delta_{\ell'-\ell,r}\left[{\rm sgn}(n-n')\right]^{\xi} \cos[\Theta_{rs}^{\ell}(n,\,n')]\right\}\ ,
\end{eqnarray}

\noindent
where $\xi=|n-n^\prime |$,
\begin{equation}
D_{n',n}^{rs(B)}=\sqrt{\frac{n_1!}{n_2!}}\,{\rm e}^{-W_{rs}/(2\beta)}\left(\frac{W_{rs}}{\beta}\right)^{\xi/2}L_{n_1}^{(\xi)}\left(\frac{W_{rs}}{\beta}\right)\ ,
\end{equation}
$n_1={\rm min}(n,\,n')$, $n_2={\rm max}(n,\,n')$, $L_n^{(m)}(x)$ is the associated Laguerre polynomial,
\begin{equation}
W_{rs}=\frac{\pi \left( s^2 K_2^2 + r^2 K_1^2 \right)}{K_1 K_2)} \ , \hspace{0.2 in} K_2=\frac{2 \pi}{d_y} \ , \hspace{0.2 in}
D_{n',n}^{rs(A)}=D_{n'-1,n-1}^{rs(B)} \ ,
\end{equation}
with $T_\ell^s$:
\begin{eqnarray}
&& T_\ell^s = \pm 2\cos\left[\frac{s (k_xd_x-2\ell\pi)}{\beta}\right] \,\, \delta_{\xi,\{4 N, \, 4 N+2\}}\ , \\
\nonumber
&& T_\ell^s = \pm 2\sin\left[\frac{s(k_xd_x-2\ell\pi)}{\beta}\right] \,\, \delta_{\xi,\{4 N+1, \, 4 N+3\}}\ ,
\end{eqnarray}
and
\begin{equation}
\Theta^\ell_{rs}(n',\,n)= \frac{s}{\beta} \left[k_xd_x-2\pi\left(\ell+r/2\right)\right] - 
{\rm sgn}(n'-n)\,\xi\tan^{-1}\left(\frac{sd_x}{rd_y}\right) \ .
\end{equation}
The magnetic band structure for this modulated system is a solution of the eigenvector problem
${\cal M} \bigotimes\vec{\cal A}(\vec{k}_{||})=0$ with the coefficient matrix $\tensor{\cal M}$
given by

\begin{equation}
\{{\cal M}\}_{j,\,j'}=\left[E^{\mu}_n-\varepsilon({\vec k}_{||})\right]
\delta_{n,n'}\delta_{\ell,\ell'}\delta^{(n)}_{\mu,\mu'} + V^{\ell',n',\mu'}_{\ell,n,\mu}(\vec{k}_{||})\ ,
\label{MM}
\end{equation}
where $\delta^{(n)}_{\mu,\mu'}=1$ for $n=0$ and $\delta^{(n)}_{\mu,\mu'}=\delta_{\mu,\mu'}$ for $n>0$, $j=\{n,\,\ell,\,\mu\}$ is the composite index, and $\{\vec{\cal A}(\vec{k}_{||})\}_j={\cal A}^\mu_{n,\ell}(\vec{k}_{||})$ is the eigenvector. The eigenvalue $\varepsilon_\nu({\vec k}_{||})$
of the system is determined by ${\rm Det}\{\tensor{\cal M}\}=0$.

\par
\medskip
\subsection{Numerical Results for Band Structure}
\begin{figure}
\centering
\includegraphics[width=0.6\textwidth,height=0.3\textwidth]{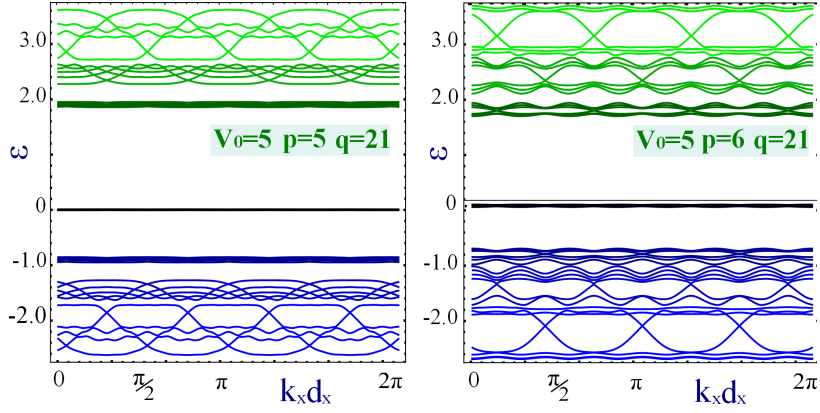}
\caption{(Color online). Energy  dispersion  as functions of $k_xd_x$ for chosen values of $V_0$ and magnetic flux
$p/q$ in units of the flux quantum. The energy is scaled in terms of $v_F\sqrt{eB\hbar}$.}
\label{FIG:1}
\end{figure}
\begin{figure}
\includegraphics[width=0.9 \textwidth,height=0.5\textwidth]{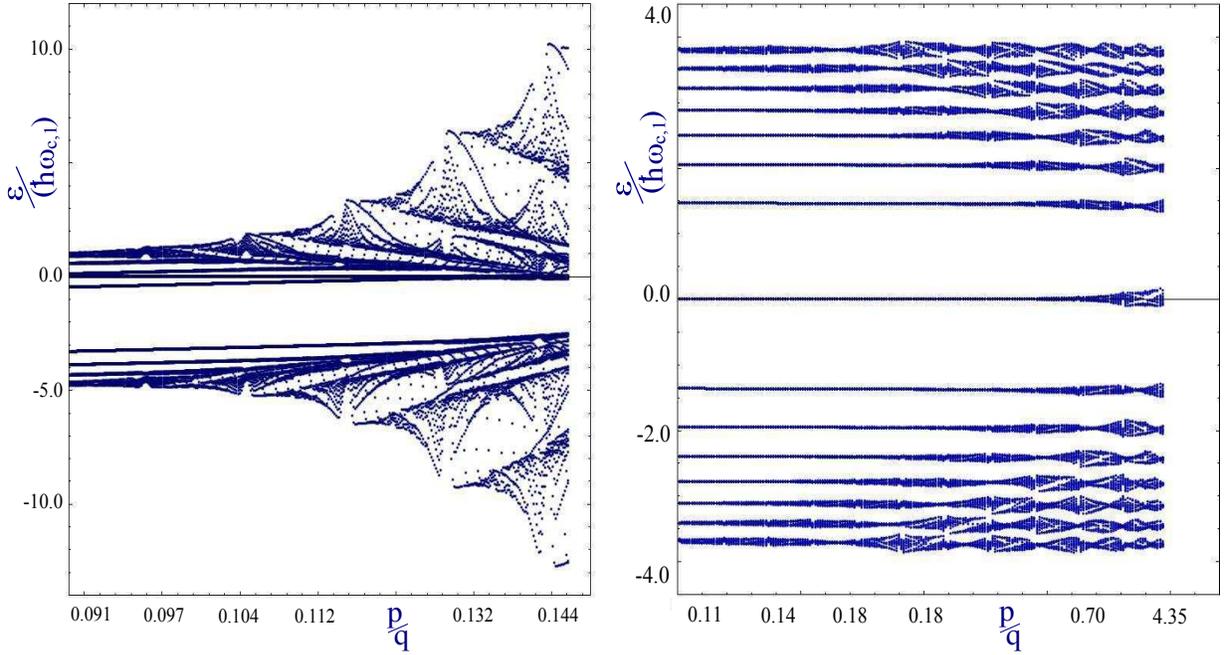}
\caption{(Color online). Energy eigennvalues as  a function of $p/q$ for graphene for chosen values when
$V_0=20$ (on the left) and $V_0=0.5$ (on the right) ,
$k_xd_x=0.3,\ k_yd_x=0.3$. The energy unit is $v_F\sqrt{eB\hbar}$.}
\label{FIG:2}
\end{figure}

\begin{figure}
\centering
\includegraphics[width=0.46 \textwidth,height=0.46\textwidth]{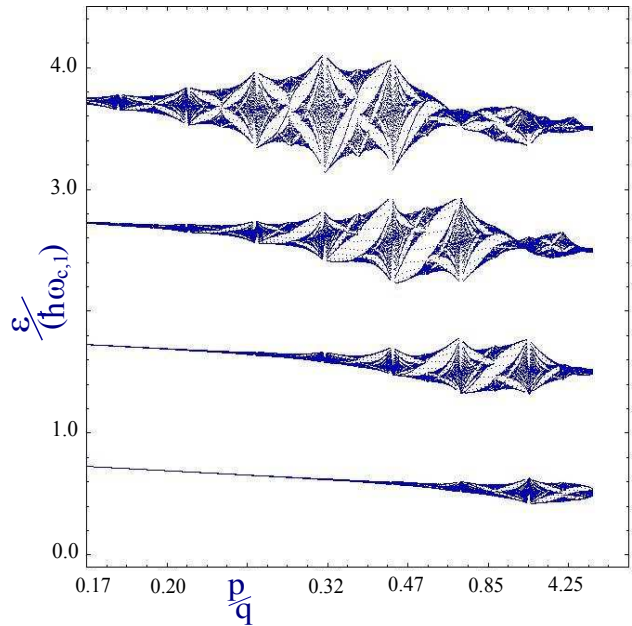}
\caption{(Color online). The four lowest Landau subbands for a weakly modulated 2DEG with chosen $V_0=0.5$
and $k_xd_x=k_yd_x=0.3$. The unit of energy  is $\hbar$  times the cyclotron frequency. }
\label{FIG:3}
\end{figure}

In Fig.\ \ref{FIG:1}, we present the dispersion curves as  a function of $k_x d_x$
for  chosen value of $V_0$ and two pairs of values of $p$ and $q$  corresponding to two
different magnetic field strengths.  In each case,  there are $p$ Landau subbands,
$q/p$  determines the number of oscillation periods in the first  Brillouin zone
for each of these subbands. Both the valence and  conduction subbands are shifted upward
but the conduction subbands are shifted more than the valence subbands for each corresponding
Landau label for the unmodulated structure.  This shift is increased when  the modulation amplitude
is increased. The original zero-energy Landau level is  on;y slightly broadened
and is the least affected by $V_0$. If the sign of the modulation amplitude is
reversed to correspond to an array of quantum dots, then the subbands are  all shifted downward.
from their positions for an unmodulated monolayer graphene.

Figure \ref{FIG:2} shows the results of our calculations for  the energy eigenvalues  of
modulated graphene  as a function of magnetic flux. We included  the $n=0,\pm 1,\pm2,
\pm3,\pm 4$ as we did in obtaining Fig.\ \ref{FIG:1}. For weak magnetic fields, the Landau
levels in both valence and conduction bands  are  slightly broadened into narrow  subbands
but shifted upward by the perturbing potential $V_0$. Another effect due to modulation
is to cause these Landau bands to have negative slope at weak magnetic fields
which then broaden enough at higher magnetic fields to produce Landau orbit mixing,
reflecting the  commensurability between the magnetic and lattice Brillouin zones.
\par
\begin{figure}
\includegraphics[width=0.98 \textwidth,height=0.46\textwidth]{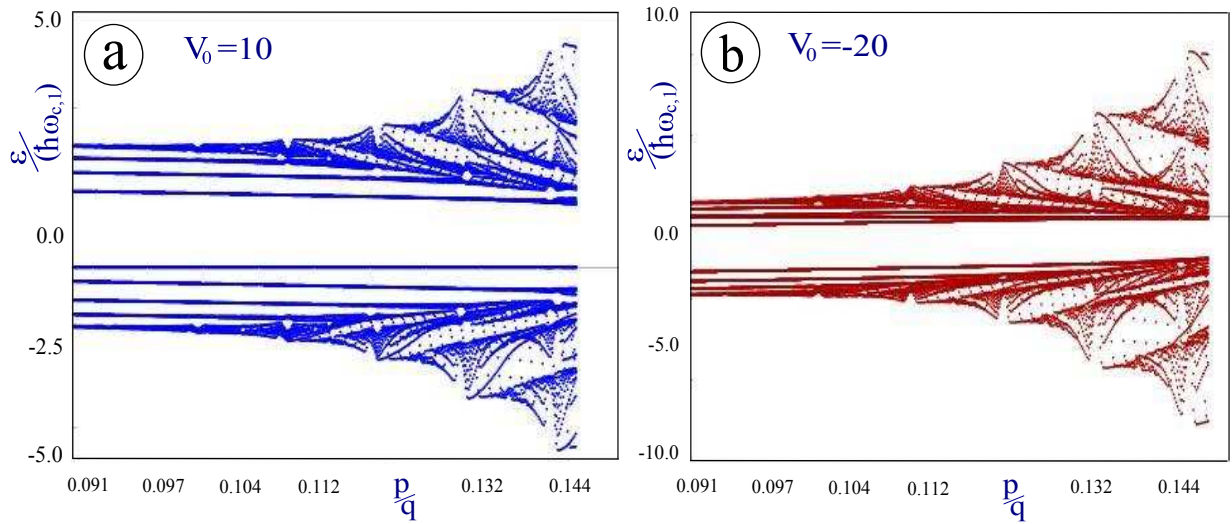}
\caption{(Color online). Energy eigennvalues as  a function of $p/q$ for graphene.
Panel $ (a)$ corresponds to $V_0=  10$ (on the left) and panel $(b)$
- to $V_0= - 20$ (negative potential), $k_xd_x=0.3,\ k_yd_x=0.3$.
 Energy is given in the units is $v_F\sqrt{eB\hbar}$.}
\label{FIG:4}
\end{figure}

The lowest perturbed Landau subband which originates from the unperturbed $n=0$ Landau level
merges with the resulting butterfly spectrum at the highest magnetic field compared to
the $n=1,2,3,4$ Landau levels in the conduction band.  The onset of the butterfly takes  place
around $p/q=1/5$ which would correspond to  a magnetic field $B\approx 2$ T for $d_x=d_y=10$ nm.
Furthermore, our calculations have shown that the symmetry between the valence and conduction
bands is destroyed by modulation. There is always mixing of the subbands regardless of the value for $V_0$.
This is in contrast with modulated 2DEG where for weak $V_0$, the Landau subbands do not overlap
as shown in Fig.\ \ref{FIG:3}. The lowest subband is shifted upward like the other
subbands but is not widened as much as the higher subbands. The feature of self-similarity is also
apparent in the excited subbands at intermediate magnetic fields. There is only a shift and broadening
of the subbands in the low and high magnetic field regimes for modulated 2DEG.

\par
\medskip
\section{Electron tunneling in the presence of magnetic field}

A goal  of this paper is to investigate single-particle properties of graphene
in the presence of both electric and magnetic fields. Consequently, to complement
our derived results in the preceding calculations,  we now turn our attention
to  the standard Klein tunneling problem   through a  square  potential
barrier \cite{Kats, mine1, mine2}
in the presence of the uniform perpendicular magnetic field.

\par
Magnetic barrier and
confinement potential for Dirac-Weyl quasiparticles in graphene were addressed in \cite{Martino}.
The reported results support the result that  it is not possible  for an  electron to tunnel   in
the presence of a uniform magnetic field. This statement follows     from the fact that
the two substantially different Landau gauge $({\bf A} = - B x \hat{e}_y)$ and the Symmetric gauge
$({\bf A} = \frac{1}{2} {\bf B} \times  {\bf r})$ are expected to result in   invariant observables such as
current density or electron momentum.

 \par
The electrostatic potential barrier is specified by

\begin{equation}
U(x) = U_0 \left[{\theta(w+x)+\theta(w-x) - 1}\right] \\
\end{equation}
and the vector potential ${\bf A}(x)$ due to the   magnetic field
${ \bf B}(x) = B_0 \Theta(d^2-x^2) \hat{e}_z$ is
has the following form
\begin{eqnarray}
&& {\bf A}(x) = A(x) \hat{e}_y  \, , \\
\nonumber
&& A(x) = \frac{1}{e \ell_B^2}  \left[ {- w \theta(w+x)+ x \theta(x^2-d^2) + w \theta(w-x)}\right] \, .
\end{eqnarray}
For this, the eigenvalue equations  the spinor wave function with components $\Psi_a (x)$ and $\Psi_b (x)$
and energy $\epsilon$  are
\begin{widetext}
\begin{eqnarray}
&& \hbar \left({\frac{\partial}{\partial x} + p_y + e A_y (x) }\right) \Psi_b (x) + i \; U(x) \Psi_a (x) = i \epsilon \Psi_a (x) \,  , 
\nonumber\\
\label{a1}
&& \hbar \left({\frac{\partial}{\partial x} - p_y - e A_y (x) }\right) \Psi_a (x) + i \; U(x) \Psi_b (x)= i \epsilon \Psi_b (x) \, .
\end{eqnarray}
\end{widetext}
where $U(x)=U_0$ is the electrostatic potential in the barrier region.
\par
First, we notice that due to the specific spatial dependence of the potentials  for
both ${{\bf A}(x)}$ and $V(x)$, the transverse component of the electron momentum is
conserved. This leads to the form   of the electron wave function in each region.
The system [\ref{a1}], determining the wave function components results in the following equations:
\begin{eqnarray}
&& \hbar^2 \frac{d^2 \psi_{a/b}(x)}{d x^2} - \left({ \pm e \frac{dA(x)}{dx}+( \hbar k_y + e A(x))}\right)\psi_{a/b}(x) = (\epsilon - U(x))^2 \psi_{a/b}(x) \ .
\end{eqnarray}
In the barrier region ,this equation has solutions in the form of parabolic cylinder functions.
The incoming particle, incident at angle $\phi$ with the normal to the barrier,  has
wave function  on the left of the barrier given by
\begin{eqnarray}
&& \Psi_{in}(x)  = \frac{1}{\sqrt{2}} \left({
\begin{array}{c}
1 \\
\frac{p_x + i\left({p_y - \hbar d/ \ell_B^2 }\right)}{\vert \mathbf{p} \vert}
\end{array}
}\right) \texttt{e}^{i \hbar p_x x} \ ,
\end{eqnarray}
where  $ p_x = \epsilon \cos \phi$, $p_y = \epsilon \sin \phi + \hbar d / \ell_B^2$
and $\texttt{s} =[\epsilon - U(x)]$. The solution within the barrier region is
given as follows
\begin{equation}
\Psi_{b}(x) = \mathcal{N} \left({
\begin{array}{c}
\mathcal{D}_{\eta-1} (\pm \sqrt{2} \zeta) \\
\texttt{s} \sqrt{1/\eta} \; \mathcal{D}_{\eta} (\pm \sqrt{2} \zeta)
\end{array}
}\right) \texttt{e}^{i \hbar p_x x}
\end{equation}
with $\zeta=x/\ell_B + p_y \ell_B$ and $\eta=((\epsilon -U(x)) \ell_B)^2 / 2$.
\par
 \begin{figure}
\centering
\includegraphics[width=0.79 \textwidth,height=0.6\textwidth]{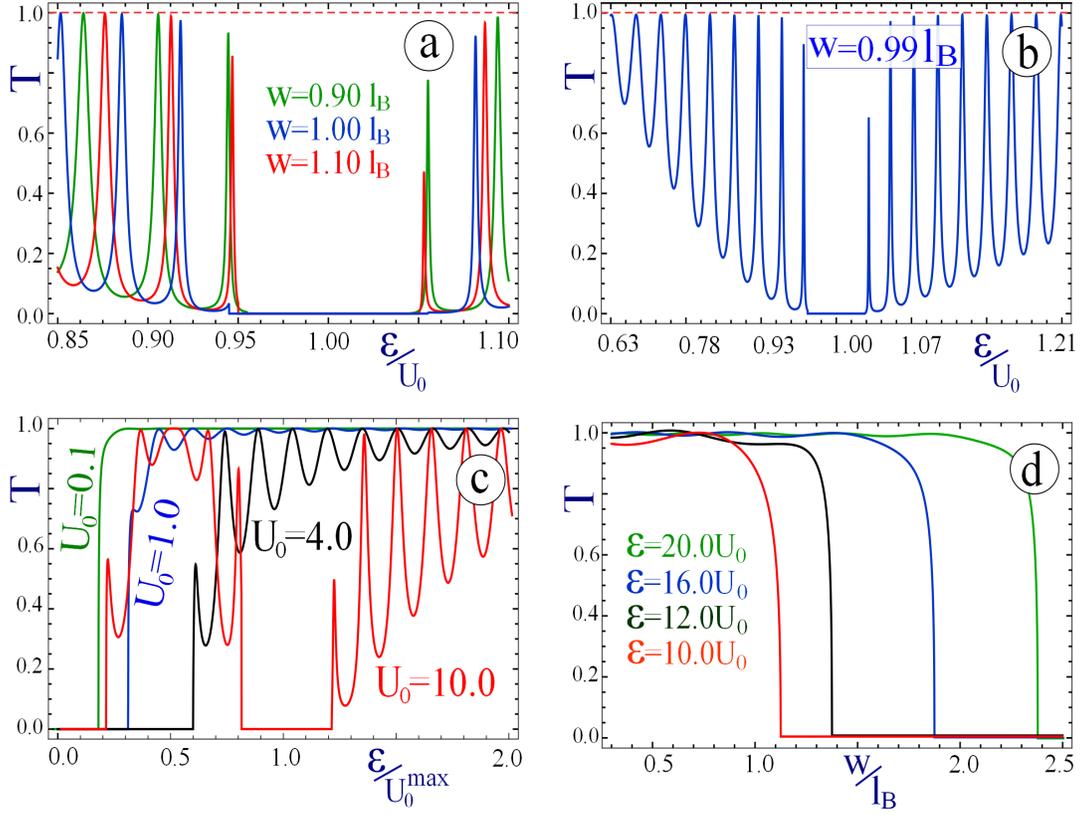}
\caption{ (Color online). Transmission probability as  functions of  the incoming electron energy
and the barrier width. Panel $(a)$ shows how the transmission depends on
the incoming electron energy in the vicinity of the electrostatic barrier height $U_0$. The
results are provided for chosen values of the barrier width: $w/\ell_B =
0.9, \, 1.0, \, 1.1$. Panel $(b)$ presents results for $w/\ell_B = 0.99$
and a wider range of energy. Panel $(c)$ shows the energy dependence of transmission
for various barrier heights $U_0$, with one of them in the vicinity of zero.
The energies are plotted in the units of the largest barrier height $U_0^{max}$.       
Plot $(d)$ shows the transmission probability as a function of    barrier
width for chosen values of the incoming particle energy $\epsilon/U_0 = 20, \,
16, \, 12$ and $10 \, $.}
\label{FIG:5}
\end{figure}
In contrast to the electrostatic potential barrier, the magnetic barrier
leads to   confinement of   Dirac electrons in graphene.
 \begin{figure}
\centering
\includegraphics[width=0.85 \textwidth,height=0.36\textwidth]{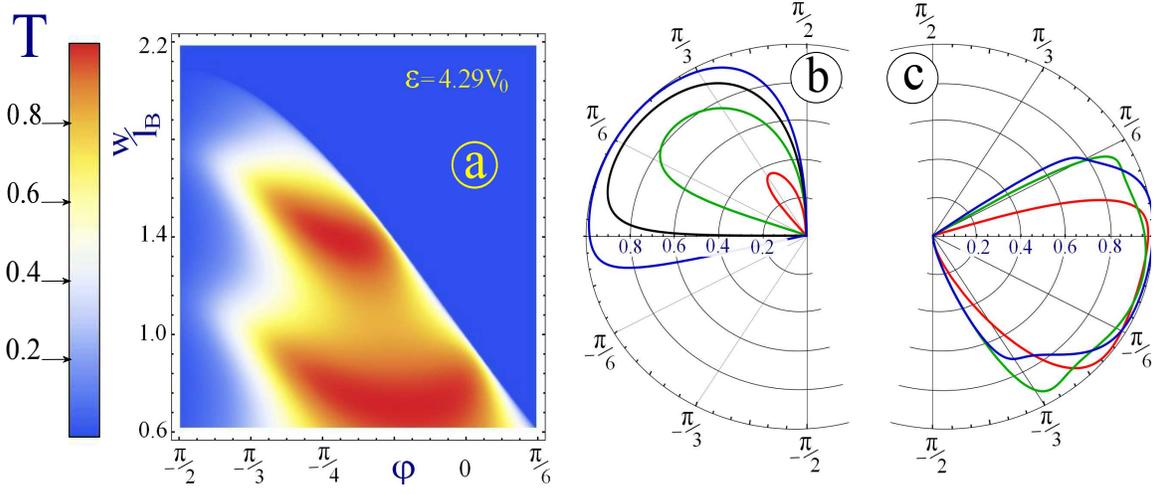}
\caption{(Color online).  Angular dependence of the electron transmission.
Panel $(a)$ is a density plot of the transmission probability as a function of  both
incoming energy and the angle of incidence. Panels $(b)$ and $(c)$  present
polar plots of electron transmission. The results in panel $(b)$ are shown
for chosen values of the incoming energy ($ \epsilon/ U_0 = 3.75, \, 5.00, \,
6.25$ and $7.5$) and in panel $(c)$   for chosen barrier width values
($w/\ell_B = 1.2, \, 1.3$ and $1.4$).}
\label{FIG:6}
\end{figure}
In agreement with    \cite{Martino}, we find  that for chosen incoming energy and
angle of incidence, there is a maximum width of the barrier for finite
transmission probability. When we also include an electrostatic potential barrier
and take into account electron-hole transition, the energy is renormalized
 in the barrier region and may be arbitrarily  small,
which makes the tunneling impossible even for low-width barriers.
 However, for the case when $U_0 \gg \epsilon$, the transmission probability
is restored  and may be equal or close to unity, depending on the
barrier width. As far as the angular dependence of the transmission
is concerned, the largest tunneling probability does not correspond  to
head-on collision $(\phi = 0)$, but at a finite angle of incidence,
 revealing the asymmetry due to  the applied magnetic
field. A sharp drop in the transmission coefficient close to the critical
 thickness of the barrier corresponds in the classical
limit to  circular trajectories of the electrons.
\par
The transmission coefficient is obtained by matching the wave function
at the boundary of each region, keeping in mind that the longitudinal momentum
does not remain constant in the barrier region, i.e.,
\begin{widetext}
\begin{eqnarray}
&& T=\frac{p_{x,3}}{p_{x,1}} \vert t \vert^2 \, ,\\
\nonumber
&& T=\frac{\sqrt{2} \texttt{e}^{- i \hbar w (p_{x,1}+p_{x,3})} \left({1+\texttt{e}^{2 i \phi}}\right) \ell_B (\epsilon - U_0)\left\{{\Lambda^{2,+} \Xi^{2,+} + \Lambda^{2,-} \Xi^{2,+} }\right\}}{2 i \texttt{e}^{i \phi} \Upsilon_1 + i \texttt{e}^{i \phi_2} (U_0 - \epsilon) \epsilon \Upsilon_2 +\sqrt{2} \texttt{e}^{i(\phi+\phi_2)} \epsilon \Upsilon_3 -\sqrt{2}(U_0 - \epsilon) \Upsilon_4} \ . \\
\nonumber
\end{eqnarray}
\end{widetext}
In this notation, $\Upsilon_1 = \Xi^{1,+} \Lambda^{2,-} - \Xi^{1,+}\Lambda^{2,-}$,
$\Upsilon_2 = \Lambda^{1,+} \Lambda^{2,-} - \Lambda^{1,-} \Lambda^{2,+}$, $\Upsilon_3 = \Lambda^{1,+}
\Xi^{2,-} + \Lambda^{2,+} \Xi^{2,+}$, $\Upsilon_4 = \Lambda^{2,+} \Xi^{1,-} + \Xi^{2,-} \Lambda^{1,+}$
and
\begin{eqnarray}
\nonumber
&& \Lambda^{1, \pm}(\epsilon, U_0, \phi, w) = \mathcal{D}_{\eta - 1}( \pm \sqrt{2} (-\frac{w}{\ell_B} + p_y \ell_B / \hbar)) \, ,\\
\nonumber
&& \Xi^{1, \pm}(\epsilon, U_0, \phi, w) = \mathcal{D}_{\eta}( \pm \sqrt{2} (-\frac{w}{\ell_B} + p_y \ell_B / \hbar)) \, ,\\
\nonumber
&& \Lambda^{2, \pm}(\epsilon, U_0, \phi, w) = \mathcal{D}_{\eta - 1}( \pm \sqrt{2} (\frac{w}{\ell_B} + p_y \ell_B / \hbar)) \, , \\
&& \Xi^{2, \pm}(\epsilon, U_0, \phi, w) = \mathcal{D}_{\eta}( \pm \sqrt{2} (\frac{w}{\ell_B} + p_y \ell_B / \hbar)) \, .
\end{eqnarray}

\par
\medskip
\par
Numerical results for  electron transmission are presented in Fig.[\ref{FIG:5}] and Fig.[
\ref{FIG:6}]. Our general conclusion is that the transmission incorporates properties of  both
electric and magnetic potential barriers. Equivalent transmission resonances may be observed
for $\epsilon < U_0$, which corresponds to electron-hole transition at the boundary of the
potential region.
\par
We note that the \textit{electrostatic potential} $U_0$ increases and drops
 sharply at the boundaries, whereas the \textit{magnetic vector potential} is continuous.
 However the electrostatic potential is uniform inside the barrier region. This significant
 difference in the spatial dependence suggests that we may consider the effect due
  to each potential separately as an adequate approximation.  As a matter of  fact, we take into
  account the refraction due to the electrostatic potential barrier first, and then deal with  the
transmission of the new state in the magnetic barrier. Consequently, the \textit{energy - width}
relationship, determining the condition of complete reflection, must now  include the energy
in the barrier region $\epsilon - U_0$. As a result, the transmission drops to zero in the
vicinity of $\epsilon = U_0$, as we can clearly see from Fig.[\ref{FIG:4}].

\par
We define the transmission resonances as asymmetric the peaks of the transmission for both
electron and hole states in the barrier region. The term comes from the theory of the
\textit{Klein paradox} and is employed  to separate the head-on complete transmission with
its peaks corresponding to the different angles of incidence. In contrast to the Klein paradox
which persists for the barriers of arbitrary width and height, the resonances occur for different
angles of incidence and the particle energies which depend  on the above mentioned parameters.
As far as the transmission resonances in the presence of magnetic field are concerned, they exhibit
similar properties to the case of pure electric barrier.
\par
We also conclude that the transmission resonances disappear in the limit when $U_0 \to 0$,
i.e., when only the magnetic barrier is present. It is interesting to note that even for a
finite  value of $U_0$, there are no peaks for $\epsilon < U_0$. It could be explained by
the fact that for  incoming particle energies close to the barrier height, the transmission is
suppressed. Analyzing the transmission probability dependence on the barrier width, we confirm that there
exists a critical width $w_{\rm cr}$ , such that for any barrier whose width exceeds this critical value,
the transmission is completely suppressed. This corresponds to a sharp drop of transmission next
to the critical  value of the barrier width. This critical value decreases with  increasing electron
energy. For $w < w_{\rm cr}$, the transmission exhibits peaks and somewhat oscillatory but not
periodic behavior.
\par
Angular dependence of the electron transmission probability, presented in Fig.[\ref{FIG:5}], also
demonstrates several interesting features. The transmission maximum no longer corresponds to
head-on collision, but is shifted to a finite angle which is determined by the incoming electron energy. Transmission dependence shows the decrease with drastically different properties, corresponding to the
larger and smaller angles compared to the angle with the largest  transmission probability.
 For a larger angle of incidence, the decrease is moderate. However, in the opposite direction,
 we observe a sharp drop. For the barrier width dependence, we once again see oscillatory behavior
(see the density plot, panel $(a)$ of  Fig.[\ref{FIG:5}]). There are also angular transmission
resonances present.

\par
\medskip
\section{Concluding Remarks}

In this paper, we Presented a formalism for calculating the energy band structure for an
electrostatically modulated single-layer of graphene in the presence of a uniform perpendicular
magnetic field.   At low and high magnetic fields, the Landau levels are broadened into
non-overlapping minibands. However, at intermediate fields, the commensurability relations
between the  modulation period and the cyclotron radius cause a mixing of the Landau orbits
and overlapping of the subbands. The resulting picture is that of a Hofstadter butterfly
for the conduction and valence bands. The onset of the self-similarity in the energy band
structure depends on the modulation amplitude.  The minibands for graphene  always overlap
in the presence of  an electrostatic modulation, unlike the 2DEG. We also obtained the
energy dispersion for  graphene. The asymmetry of these curves in the valence and  conduction
subbands is due to modulation as may be seen from our eigenvalue equation
${\rm Det}\{\tensor{\cal M}\}=0$ where the matrix  $\tensor{\cal M}$ is defined in
Eq.\ (\ref{MM}).
\par
Also, we addressed the electron tunneling problem in the presence of both electrostatic
and magnetic potential barriers within the same region. We have found that  transmission is
suppressed for  incoming electron energies close to the barrier height and for
barrier widths exceeding  critical values. In contrast, transmission resonances
exist for the case of electron-hole transmission at the boundary of the potential region.
The angular dependence of the electron transmission probability demonstrates the shift
of the transmission maximum to a  finite angle of incidence, compared to Klein paradox for
purely electrostatic potential barriers.


\acknowledgments
This research was supported by  contract \# FA 9453-07-C-0207 of AFRL.


\begin{thebibliography}{9}
	
\bibitem{azbel} M. Ya. Azbel, Sov. Phys. JETP, {\bf 19}, 634 (1964).

\bibitem{hofstadter}D. R. Hofstadter,
   Phys. Rev.B  {\bf 14}, 2239 (1976).

\bibitem{GG}   Godfrey Gumbs and Paula Fekete,
    Phys. Rev. B {\bf 56}, 3787 (1997).

\bibitem{wannier}
    G. H. Wannier, Phys. Stat. Sol. b {\bf 100}, 163 (1980).

\bibitem{rauh} A. Rauh, G. H. Wannier, and G. Obermair,
Phys. Stat. Solidi b {\bf 63}, 215 (1974).

\bibitem{rauh2}  A. Rauh, Phys. Stat. Sol. b {\bf 69},
K9 (1975).

\bibitem{rauh3}  H. W. Neumann and A. Rauh, Phys. Stat. Solidi b {\bf 96}, 233 (1979).

\bibitem{kohmoto1}Y. Hasegawa, Y. Hatsugai, M. Kohmoto, and G. Montambaux,
Phys. Rev. B {\bf 41}, 9174 (1990).

\bibitem{kohmoto2}Y. Hatsugai and M. Kohmoto, Phys. Rev. B {\bf 42},
   8282 (1990).

\bibitem{pfannkuche}D. Pfannkuche and R. R. Gerhardts,
   Phys. Rev. B {\bf 46}, 12606 (1992).

\bibitem{pfannkuche2}D. Pfannkuche and R. R. Gerhardts,
    Surf. Sci. {\bf 263}, 324 (1992).

\bibitem{thouless}D. J. Thouless, Phys. Rev. B {\bf 28}, 4272 (1983).

\bibitem{claro}F. Claro, Phys. Stat. Sol. (b) {\bf 104}, K31 (1981).

\bibitem{schellnhuber}H. J. Schellnhuber and G. M. Obermair, Phys.
Rev. Lett. {\bf 45}, 276 (1980).

\bibitem{perschel}T. Perschel and T. Geisel. Phys. Rev. Lett.
{\bf 71}, 239 (1993).

\bibitem{wu}X. Wu and S. E. Ulloa, Phys. Rev. B {\bf 47}, 10028
(1993).

\bibitem{silbernauer}H. Silbernauer, J. Phys.: Condens. Matter {\bf 4}, 7355 (1992).

\bibitem{sokoloff}J, B. Sokoloff, Physics Reports {\bf 126}, 189 (1985).

\bibitem{cui1}O. K\"uhn, V. Fessatidis, H. L. Cui, P. E. Selbmann, and N.Horing,
Phys. Rev.B {\bf 47}, 19, 13019 (1993).

\bibitem{bilayer}    N. Nemec and G. Cunibert,
Phys.Rev.B  {\bf 75}, 201404(R) (2007).

\bibitem{ye} P. D. Ye, D. Weiss, R. R. Gerhardts, M. Seeger, K. von Klitzing,
K. Eberl, and H. Nickel, Phys. Rev. Lett. {\bf 74}, 3013 (1995).

\bibitem{nat1} C. R. Dean et al., Nature {\bf 497}, 598 - 602 (30 May 2013).

\bibitem{nat2} L. A. Ponomarenko et al., Nature {\bf 497}, 594 - 597 (30 May 2013). 


\bibitem{petters} M. Zarenia, P. Vasilopoulos, and F. M. Peeters, Phys.Rev. B,
{\bf 85}, 245426 (2012).

\bibitem{Kats} M. I.  Katsnelson, K. S. Novoselov, and A. K. Geim, Nature Physics, {\bf 2}, 620 (2006).

\bibitem{mine1} Andrii Iurov, Godfrey Gumbs, Oleksiy Roslyak and  Danhong Huang,
Journal of Physics.:  Condensed Matter, {\bf 24}, 1, 015303 (2012)

\bibitem{mine2}Andrii Iurov, Godfrey Gumbs, Oleksiy Roslyak and  Danhong Huang,
 Journal of Physics.: Condensed Matter, {\bf 25}, 13, 135502 (2013)

\bibitem{Martino} A. De Martino, L. Dell'Anna and R.Egger,
Solid State Communications, {\bf 144}, 12 (2007).

\end{thebibliography}
\end{document}